# Instantaneous GNSS attitude determination: A Monte Carlo sampling approach

Xiucong Sun, Chao Han, Pei Chen *

*School of Astronautics, Beihang University, Beijing 100191, China*

**Abstract** A novel instantaneous GNSS ambiguity resolution approach which makes use of only single-frequency carrier phase measurements for ultra-short baseline attitude determination is proposed. The Monte Carlo sampling method is employed to obtain the probability density function of ambiguities from a quaternion-based GNSS-attitude model and the LAMBDA method strengthened with a screening mechanism is then utilized to fix the integer values. Experimental results show that 100% success rate could be achieved for ultra-short baselines.

## 1. Introduction

The Global Navigation Satellite System (GNSS) is conceived as a viable alternative or complement to traditional attitude sensors for attitude determination of land, sea, air, and space vehicles [1-14]. The key to high-precision GNSS attitude determination is integer ambiguity resolution. Compared with motion-based ambiguity resolution methods [11,15-17] which exploit the time-varying receiver-satellite geometry, search-based methods [7,12-14,18-25] achieve instantaneous attitude determination and can be used in GNSS-challenged environments where frequent losses of lock occur. Among various search-based methods, the LAMBDA (Least-squares AMBiguity Decorrelation Adjustment) method [20,26] and its variants [6-9,21,22,27-28] have been widely used for their numerical efficiency and high success rate. The LAMBDA method requires both code and carrier phase observations. The contribution of code measurements is that they restrict the search space of integer ambiguities from infinity to a small region. A high success rate can benefit from a low code noise level. Conversely, a poor quality of code measurements results in degraded performance [8].

The present study seeks a solution to LAMBDA ambiguity resolution for attitude determination without aid of code measurements. This will be useful for applications under higher code noise and/or multipath environments

* Corresponding author. Tel.: +86 10 82316535.
E-mail address: chenpei@buaa.edu.cn (P. Chen).

(low-end GNSS receivers, urban canyons, etc.). The multivariate GNSS-attitude model [8,9,22] associates double-differenced (DD) phase ambiguities with multi-antenna platform attitude in a probabilistic manner. The scale of the probability space of ambiguities is related to baseline length. For ultra-short baseline attitude determination, the ambiguity probability space is sufficiently small. There is no need to use additional code measurements to restrict the ambiguity space. In this study, the Monte Carlo sampling (MCS) method [29-32] is employed to construct the probability density function (pdf) of ambiguities conditioned on DD phase observations. The constructed pdf is used to obtain the expectation and covariance of ambiguities, which are fed to standard LAMBDA for integer ambiguity resolution.

## 2. Quaternion-based GNSS-attitude model

Consider a set of $m$ + 1 ($m \geq 2$) receivers/antennas tracking $n$ + 1 common GNSS satellites. The doubled-differenced carrier phase observations formed with $m$ independent baselines are cast into the following multivariate GNSS-attitude model [22]

$$\begin{gathered} \boldsymbol{\Phi} = \boldsymbol{GRF} + \lambda \boldsymbol{Z} + \boldsymbol{V}; \quad \operatorname{Cov}\left(\operatorname{vec}\left(\boldsymbol{V}\right)\right) = \boldsymbol{Q} \\ \boldsymbol{R} \in \mathbb{O}^{3\times3}; \quad \boldsymbol{Z} \in \mathbb{Z}^{n\times m} \end{gathered} \tag{1}$$

where $\boldsymbol{\Phi}$ is the ($n \times m$) matrix containing the DD phase observations from the $m$ baselines, $\boldsymbol{F}$ is the ($3 \times m$) matrix of baseline coordinates (expressed in the local coordinate system of the multi-antenna platform), $\boldsymbol{R}$ is the ($3 \times 3$) matrix of coordinate transformation from the local coordinate system to the GNSS system reference system, $\boldsymbol{G}$ is the ($n \times 3$) matrix of double-differenced unit line-of-sight vectors, $\boldsymbol{Z}$ is the ($n \times m$) matrix of integer ambiguities (expressed in cycles), $\lambda$ is the carrier phase wavelength, and $\boldsymbol{V}$ is the ($n \times m$) matrix of observation noises. The vector operator vec(·) is defined as stacking the columns of the ($n \times m$) matrix $\boldsymbol{V}$ into the vector vec($\boldsymbol{V}$) of order $nm$. $\boldsymbol{Q}$ is the covariance matrix of vec($\boldsymbol{V}$) and is constructed as follows

$$\boldsymbol{Q} = \sigma^2 \begin{bmatrix} 1 & 0.5 & \cdots & 0.5 \\ 0.5 & 1 & \cdots & \vdots \\ \vdots & \vdots & \ddots & 0.5 \\ 0.5 & \cdots & 0.5 & 1 \end{bmatrix}_{m\times m} \otimes \begin{bmatrix} 4 & 2 & \cdots & 2 \\ 2 & 4 & \cdots & \vdots \\ \vdots & \vdots & \ddots & 2 \\ 2 & \cdots & 2 & 4 \end{bmatrix}_{n\times n} \tag{2}$$

where $\sigma$ is the standard deviation of undifferenced phase noise and the operator $\otimes$ denotes the Kronecker product. The unknowns to be resolved in model (1) are the integer ambiguities and the $3 \times 3$ orthonormal attitude matrix $\boldsymbol{R}$.

In this study, the attitude matrix is parameterized with quaternion representation. Model (1) is further written as the following quaternion-based observation equation

$$\begin{gathered}\boldsymbol{\Phi} = \boldsymbol{G}\boldsymbol{R}(\boldsymbol{q})\boldsymbol{F} + \lambda\boldsymbol{Z} + \boldsymbol{V}; \quad \mathrm{Cov}\left(\mathrm{vec}(\boldsymbol{V})\right) = \boldsymbol{Q} \\ \boldsymbol{q} \in \mathbb{R}^{4\times 1}; \quad \boldsymbol{Z} \in \mathbb{Z}^{n\times m}\end{gathered} \tag{3}$$

with

$$\boldsymbol{R}(\boldsymbol{q}) = \begin{bmatrix} q_1^2 - q_2^2 - q_3^2 + q_4^2 & 2(q_1q_2 + q_3q_4) & 2(q_1q_3 - q_2q_4) \\ 2(q_1q_2 - q_3q_4) & -q_1^2 + q_2^2 - q_3^2 + q_4^2 & 2(q_2q_3 + q_1q_4) \\ 2(q_1q_3 + q_2q_4) & 2(q_2q_3 - q_1q_4) & -q_1^2 - q_2^2 + q_3^2 + q_4^2 \end{bmatrix} \tag{4}$$

where the quaternion $\boldsymbol{q} = (\boldsymbol{q}_0, q_4)^T$, $\boldsymbol{q}_0 = (q_1, q_2, q_3)^T$, and $\boldsymbol{q}^T\boldsymbol{q} = 1$.

## 3. Monte Carlo sampling approach

### *3.1. Constructing pdf of ambiguities*

Based on model (3), the DD ambiguities are functions of attitude quaternion and DD phase observations

$$\boldsymbol{Z} = \frac{1}{\lambda}\left(\boldsymbol{\Phi} - \boldsymbol{G}\boldsymbol{R}(\boldsymbol{q})\boldsymbol{F} - \boldsymbol{V}\right) \tag{5}$$

where the quaternion $\boldsymbol{q}$ and observation noise matrix $\boldsymbol{V}$ are regarded as independent random variables.

The observation noises are assumed to be normally (Gaussian) distributed with mean zero and covariance $\boldsymbol{Q}$. The pdf of vec($\boldsymbol{V}$) is

$$p_{\mathrm{vec}(V)}(\boldsymbol{v}) = \mathcal{N}(\boldsymbol{v}; \boldsymbol{0}, \boldsymbol{Q}) \triangleq \frac{1}{\sqrt{2\pi \det \boldsymbol{Q}}} e^{-\frac{1}{2}\boldsymbol{v}^T\boldsymbol{Q}^{-1}\boldsymbol{v}} \tag{6}$$

The probability distribution of $\boldsymbol{q}$ can be inferred from a priori attitude information, which is usually obtained from coarse attitude sensors or dead reckoning. Given a priori quaternion with lower bound $\boldsymbol{q}^l$ and upper bound $\boldsymbol{q}^u$, a uniform distribution incorporating the norm 1 constraint can be formulated as follows

$$p_q(\boldsymbol{q}) = \mathcal{U}(\boldsymbol{q}; \boldsymbol{q}^l, \boldsymbol{q}^u) \triangleq \begin{cases} c & \boldsymbol{q}_0 \in \left[\boldsymbol{q}_0^l, \boldsymbol{q}_0^u\right],\ \|\boldsymbol{q}_0\| \le 1,\ q_4 = \pm\sqrt{1 - \|\boldsymbol{q}_0\|^2} \\ 0 & \text{elsewhere} \end{cases} \tag{7}$$

with

$$\int\limits_{\substack{\boldsymbol{q}_0\in\left[\boldsymbol{q}_0^l,\,\boldsymbol{q}_0^u\right],\\ \|\boldsymbol{q}_0\|\le 1}} c\,d\boldsymbol{q}_0 = 1 \tag{8}$$

where $\boldsymbol{q}_0^l$ and $\boldsymbol{q}_0^u$ are the vectorial parts of $\boldsymbol{q}^l$ and $\boldsymbol{q}^u$, respectively. The lower bound $\boldsymbol{q}^l$ and upper bound $\boldsymbol{q}^u$ can be set to the $3\sigma$ error bounds of a priori attitude estimation. In the absence of a priori attitude information, $\boldsymbol{q}_0^l$ and $\boldsymbol{q}_0^u$ can be set to $[-1,-1,-1]^T$ and $[1,1,1]^T$, respectively.

The pdf of ambiguities in terms of pdfs of $\boldsymbol{q}$ and vec($\boldsymbol{V}$) is [33]

$$p_{\mathrm{vec}(\boldsymbol{Z})}(\boldsymbol{z}) = \frac{\partial}{\partial \mathrm{vec}(\boldsymbol{Z})} \iint\limits_{\boldsymbol{q},\boldsymbol{V}\in D_z} p_{\boldsymbol{q}}(\boldsymbol{q})\, p_{\mathrm{vec}(\boldsymbol{V})}(\boldsymbol{v})\, d\boldsymbol{q}\, d\mathrm{vec}(\boldsymbol{V}) \tag{9}$$

with

$$D_z = \left\{\boldsymbol{q},\boldsymbol{V};\frac{1}{\lambda}\mathrm{vec}\left(\boldsymbol{\Phi}-\boldsymbol{G}\boldsymbol{R}(\boldsymbol{q})\boldsymbol{F}-\boldsymbol{V}\right)\le \boldsymbol{z}\right\} \tag{10}$$

It is difficult to obtain the analytical expression of $p_{\mathrm{vec}(\boldsymbol{Z})}(\boldsymbol{z})$. The Monte Carlo method provides a convenient alternative to represent pdfs of random variables for nonlinear/non-Gaussian systems by a set of particles with associated weights.

The procedure of constructing $p_{\mathrm{vec}(\boldsymbol{Z})}(\boldsymbol{z})$ using the Monte Carlo method is given as follows.

First, draw $N_s$ particles from $p_{\boldsymbol{q}}(\boldsymbol{q})$ and $p_{\mathrm{vec}(\boldsymbol{V})}(\boldsymbol{v})$

$$\boldsymbol{q}_i \sim p_{\boldsymbol{q}}(\boldsymbol{q}); \quad \boldsymbol{v}_i \sim p_{\mathrm{vec}(\boldsymbol{V})}(\boldsymbol{v}) \tag{11}$$

Assign the particles with weights

$$w_i = \frac{1}{N_s},\ i = 1, 2, \ldots, N_s \tag{12}$$

Second, propagate the particles according to Eq. (5)

$$\boldsymbol{Z}_i = \frac{1}{\lambda}\left(\boldsymbol{\Phi}-\boldsymbol{G}\boldsymbol{R}(\boldsymbol{q}_i)\boldsymbol{F}-\boldsymbol{V}_i\right) \tag{13}$$

Finally, approximate the pdf $p_{\mathrm{vec}(\boldsymbol{Z})}(\boldsymbol{z})$ using the particles of ambiguities $\{\boldsymbol{z}_i, w_i\}_{i=1}^{N_s}$

$$p_{\mathrm{vec}(\boldsymbol{Z})}(\boldsymbol{z}) \approx \sum_{i=1}^{N_s} w_i \delta(\boldsymbol{z}-\boldsymbol{z}_i) \tag{14}$$

where $\delta(\cdot)$ is the Dirac delta function.

*3.2. Ambiguity resolution and attitude determination*

The expectation and covariance of ambiguities can be evaluated from the particles

$$\bar{z} = \sum_{i=1}^{N_s} w_i z_i \tag{15}$$

$$\boldsymbol{P} = \sum_{i=1}^{N_s} w_i \left(z_i - \bar{z}\right)\left(z_i - \bar{z}\right)^T \tag{16}$$

The expectation and covariance are then fed to LAMBDA for integer ambiguity resolution. The LAMBDA method utilizes Z-transformation to efficiently search for the 'best' integer vector which is closest to $\bar{z}$ in the metric of $\boldsymbol{P}$ [26]. This criterion is generally valid for Gaussian distributed ambiguities. However, the pdf $p_{\mathrm{vec}(\boldsymbol{Z})}\left(z\right)$ in this study is non-Gaussian. The expectation does not contain much information about the correct ambiguities and only represents an average value evaluated on the probability space.

To address this problem, the LAMBDA algorithm implemented in this study does not return only one integer vector but a specific number of candidates in order of distance from $\bar{z}$. A screening mechanism based on internal consistency checking is employed to select the candidate which best fits the GNSS-attitude model. The screening algorithm is presented as follows.

*1. Attitude matrix extraction*

For each integer ambiguity candidate $\hat{z}_k, k = 1, 2, \ldots, N_c$, the corresponding attitude matrix $\hat{\boldsymbol{R}}_k$ can be extracted by solving the following linear matrix equation

$$\boldsymbol{\Phi} - \lambda \hat{\boldsymbol{Z}}_k = \boldsymbol{G}\hat{\boldsymbol{R}}_k \boldsymbol{F} \tag{17}$$

where $\hat{\boldsymbol{Z}}_k$ is the matrix form of $\hat{z}_k$. For the two baseline case ($m$ = 2), the solution is a 3 × 2 matrix. The third column can be computed by the cross product of the first two columns.

*2. Quaternion inversion*

The quaternion candidate $\hat{\boldsymbol{q}}_k$ can be obtained from $\hat{\boldsymbol{R}}_k$ according to

$$\boldsymbol{q}_0 = \frac{1}{4q_4}\begin{bmatrix} R_{23} - R_{32} \\ R_{31} - R_{13} \\ R_{12} - R_{21} \end{bmatrix}$$

$$q_4 = \frac{1}{2}\left(\mathrm{tr}\boldsymbol{R} + 1\right)^{\frac{1}{2}} \tag{18}$$

where $\{R_{ij}, i, j = 1, 2, 3\}$ are the elements of $\boldsymbol{R}$.

*3. Residual checking*

The quaternion and ambiguity candidates are jointly verified based on the GNSS-attitude model (3). The best candidate combination is

$$\left[\hat{\boldsymbol{q}}, \hat{\boldsymbol{Z}}\right] = \arg \min_{k=1,\ldots,N_c} \left\| \operatorname{vec}\left(\boldsymbol{\Phi} - \boldsymbol{G}\boldsymbol{R}\left(\hat{\boldsymbol{q}}_k\right)\boldsymbol{F} - \lambda \hat{\boldsymbol{Z}}_k\right) \right\| \tag{19}$$

Thus, the attitude and integer ambiguities are simultaneously determined.

The overall procedure of the MSC approach for instantaneous GNSS attitude determination is summarized in Table 1. The particles construct the probability space of ambiguities. The standard LAMBDA efficiently search the space and gives integer ambiguity candidates. The screening mechanism selects the best candidate and outputs the associated attitude quaternion simultaneously.

**Table 1**. Step-by-step overview of the Monte Carlo sampling approach for instantaneous GNSS ambiguity resolution and attitude determination.

| |
|---|
| Draw particles from a priori attitude information and observation noise distribution<br>$\boldsymbol{q}_i \sim p_{\boldsymbol{q}}(\boldsymbol{q}), \boldsymbol{v}_i \sim p_{\mathrm{vec}(\boldsymbol{V})}(\boldsymbol{v}), w_i = 1/N_s, i = 1, 2, \ldots, N_s$ |
| ↓ |
| Propagate particles and construct pdf of ambiguities<br>$\{\boldsymbol{z}_i, w_i\}_{i=1}^{N_s} \rightarrow p_{\mathrm{vec}(\boldsymbol{Z})}(\boldsymbol{z}), \bar{\boldsymbol{z}}, \boldsymbol{P}$ |
| ↓ |
| Search for ambiguity candidates using LAMBDA<br>$\bar{\boldsymbol{z}}, \boldsymbol{P} \rightarrow \hat{\boldsymbol{z}}_k, k = 1, 2, \ldots, N_c$ |
| ↓ |
| Extract attitude matrix and inverse quaternion<br>$\hat{\boldsymbol{R}}_k = \mathrm{pinv}(\boldsymbol{G}) \cdot \left(\boldsymbol{\Phi} - \lambda \hat{\boldsymbol{Z}}_k\right) \cdot \mathrm{pinv}(\boldsymbol{F}), \hat{\boldsymbol{R}}_k \rightarrow \hat{\boldsymbol{q}}_k$ |
| ↓ |
| Screen candidates and output the best quaternion and ambiguity candidate<br>$\left[\hat{\boldsymbol{q}}, \hat{\boldsymbol{Z}}\right] = \arg \min_{k=1,\ldots,N_c} \left\| \operatorname{vec}\left(\boldsymbol{\Phi} - \boldsymbol{G}\boldsymbol{R}\left(\hat{\boldsymbol{q}}_k\right)\boldsymbol{F} - \lambda \hat{\boldsymbol{Z}}_k\right) \right\|$ |

The success rate of the MCS approach is closely related to the number of particles as well as the number of ambiguity candidates. The increase in the number of particles improves the approximation accuracy of pdf of ambiguities, and the increase in the number of ambiguity candidates leads to increased opportunities for the correct ambiguities. However, the computational efficiency is degraded in the meantime. The minimum numbers of particles and ambiguity candidates may vary from case to case. Empirical values should be carefully examined for practical application. In addition, the success rate is also influenced by the baseline length. A higher success rate benefits from shorter baseline length which better restricts the ambiguity probability space.

## 4. Experimental results

### *4.1. Static test*

Three antennas connected to three 12-channel GNSS M300 receivers (manufactured by ComNav Technology of China) were mounted on a rigid static platform to collect BeiDou data with a sampling interval of 1 s (see Fig. 1). The nominal noise levels of code and phase measurements are 10 cm and 0.5 mm, respectively. The master antenna *M* and the first auxiliary antenna *A1* constitute the first baseline with a fixed length of 0.512 m. *M* and the second auxiliary antenna *A2* constitute the second baseline with a fixed length of 0.542 m. The local baseline coordinates are

$$\boldsymbol{F} = \begin{bmatrix} 0.512 & 0.223 \\ 0 & 0.494 \\ 0 & 0 \end{bmatrix} \tag{20}$$

A 1100 s observation data set were collected and analyzed offline. By postprocessing data, the correct integer ambiguities at each epoch are easily reconstructed to assess the performance of the MCS approach for ambiguity resolution.

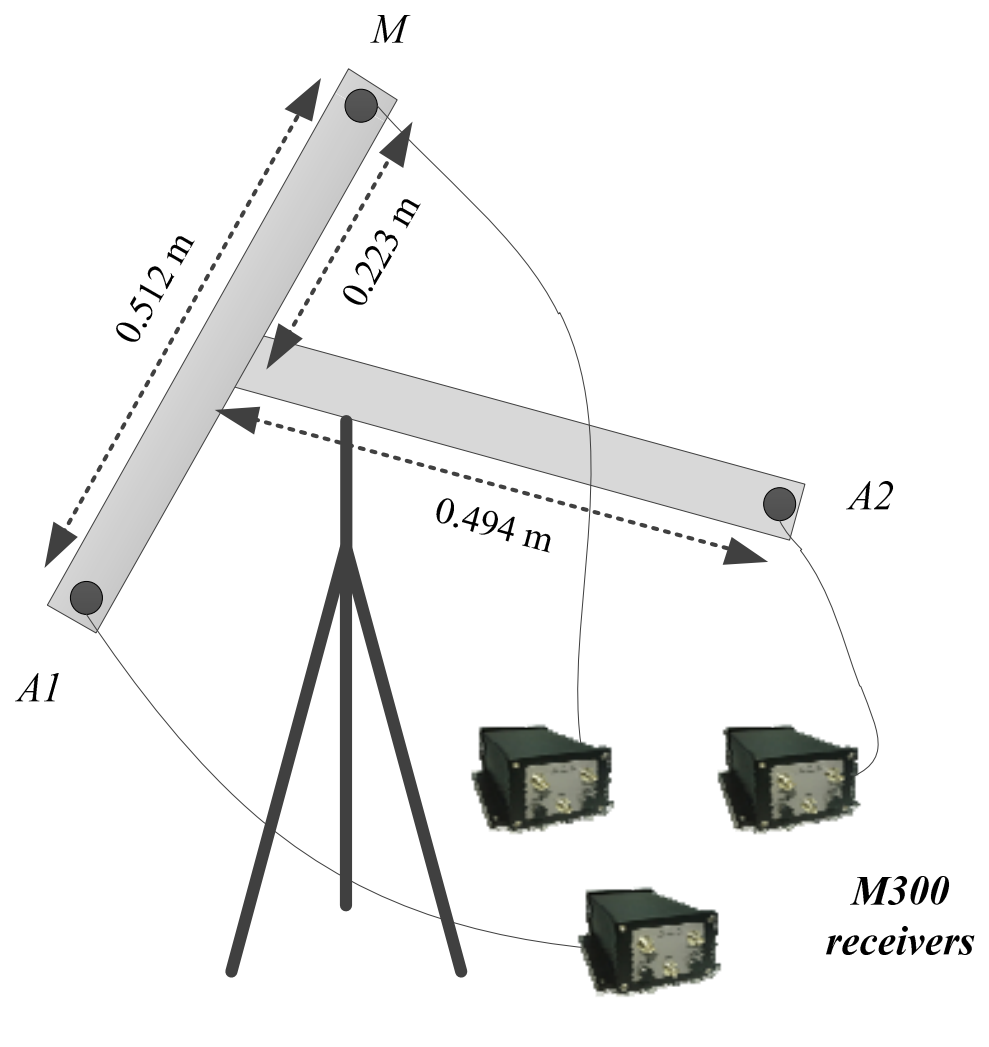

Fig. 1. Static BeiDou test: geometrical arrangement of the three antennas (location: Beijing)

In this experiment, 10 BeiDou satellites including 5 GEOs, 3 IGSOs, and 2 MEOs were commonly tracked by the antennas. Three different uncertainty levels of a priori attitude information in the form of quternion bound are tested, which are ± 0.2, ± 0.5, and ± 1, respectively. The quaternion bounds ± 0.2 and ± 0.5 approximately correspond to Euler angle errors of 20˚ and 180˚, whereas the quaternion bound ± 1 represents the case where no a priori attitude information is given. The pdf of ambiguities have been constructed by using 100, 500, and 1000 particles, respectively. The Mersenne Twister algorithm is employed to generate random particles for $\boldsymbol{q}_i$ and $\boldsymbol{v}_i$. The stochastic model (5) is used to propagate the particles to construct propability distribution of ambiguities. Figure 2 illustrates the particle cloud (1000 particles) which represents the joint distribution of the first three components of vec($\boldsymbol{V}$). It is seen that the ambiguities are not Gaussian distributed and are highly correlated.

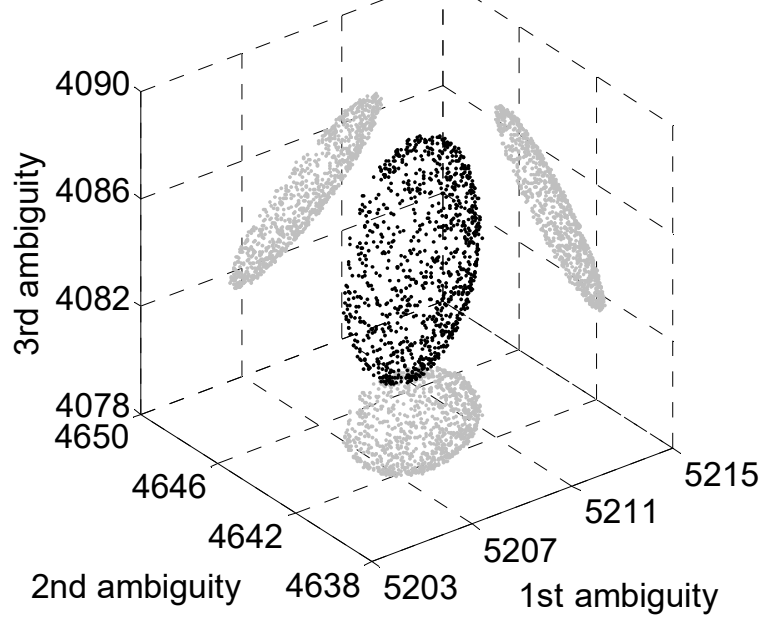

(a) Quaternion bound ± 1

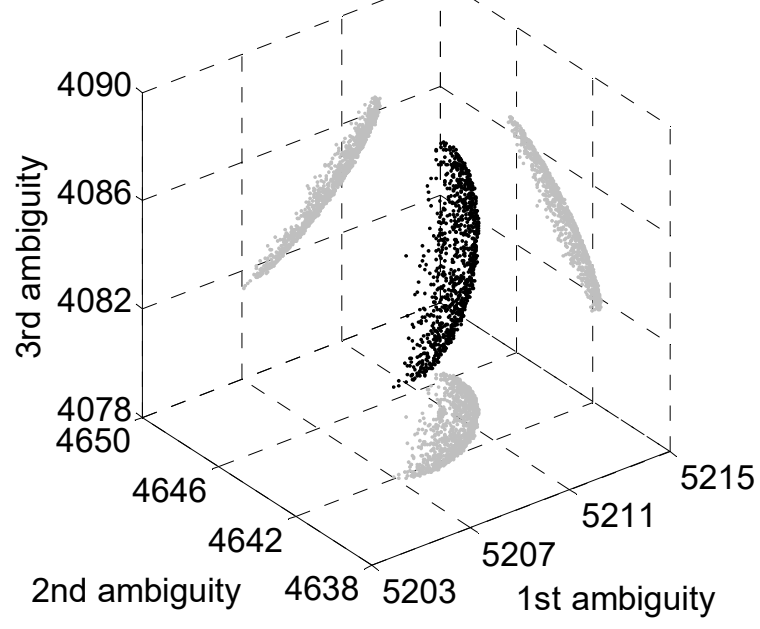

(b) Quaternion bound ± 0.5

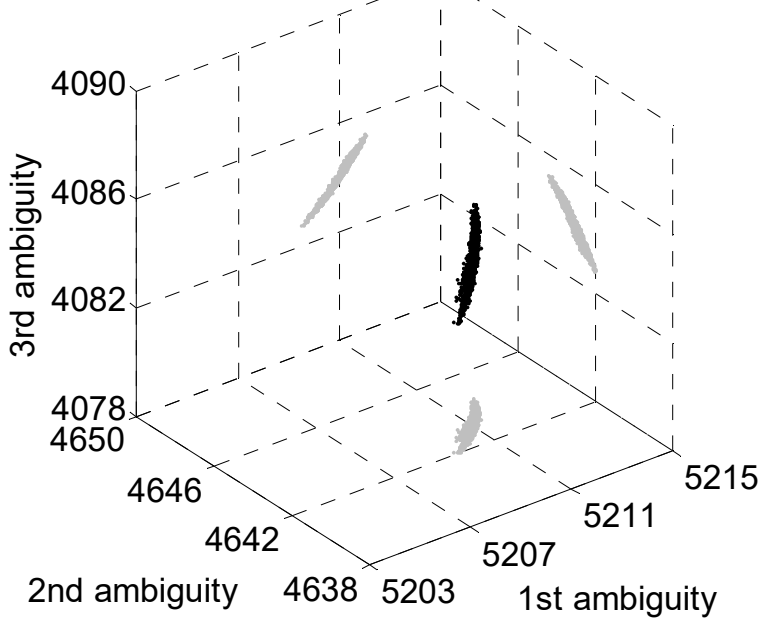

(c) Quaternion bound ± 0.2

Fig. 2. Particle cloud as representation of pdf of ambiguities.

**Table 2**. Success rate (%) for the Monte Carlo sampling approach, per number of available satellites ( $N_B$ ), number of particles ( $N_s$ ), number of integer candidates ( $N_c$ ), quaternion bounds of a priori attitude information. Rates of 100% are bolded.

| | | $N_c=5$ | | | $N_c=10$ | | | $N_c=15$ | | |
|---|---|---|---|---|---|---|---|---|---|---|
| Quaternion bound | | ± 1 | ± 0.5 | ± 0.2 | ± 1 | ± 0.5 | ± 0.2 | ± 1 | ± 0.5 | ± 0.2 |
| | $N_B=4$ | 0 | **100** | **100** | 0 | **100** | **100** | **100** | **100** | **100** |
| | $N_B=5$ | 98.9 | **100** | **100** | **100** | **100** | **100** | **100** | **100** | **100** |
| | $N_B=6$ | 99.0 | **100** | **100** | **100** | **100** | **100** | **100** | **100** | **100** |
| $N_s=100$ | $N_B=7$ | 99.9 | **100** | **100** | **100** | **100** | **100** | **100** | **100** | **100** |
| | $N_B=8$ | **100** | **100** | **100** | **100** | **100** | **100** | **100** | **100** | **100** |
| | $N_B=9$ | **100** | **100** | **100** | **100** | **100** | **100** | **100** | **100** | **100** |
| | $N_B=10$ | **100** | **100** | **100** | **100** | **100** | **100** | **100** | **100** | **100** |
| | $N_B=4$ | 0 | **100** | **100** | 5.7 | **100** | **100** | **100** | **100** | **100** |
| | $N_B=5$ | 99.3 | **100** | **100** | **100** | **100** | **100** | **100** | **100** | **100** |
| | $N_B=6$ | 99.5 | **100** | **100** | **100** | **100** | **100** | **100** | **100** | **100** |
| $N_s=500$ | $N_B=7$ | **100** | **100** | **100** | **100** | **100** | **100** | **100** | **100** | **100** |
| | $N_B=8$ | **100** | **100** | **100** | **100** | **100** | **100** | **100** | **100** | **100** |
| | $N_B=9$ | **100** | **100** | **100** | **100** | **100** | **100** | **100** | **100** | **100** |
| | $N_B=10$ | **100** | **100** | **100** | **100** | **100** | **100** | **100** | **100** | **100** |
| | $N_B=4$ | 0 | **100** | **100** | 55.9 | **100** | **100** | **100** | **100** | **100** |
| | $N_B=5$ | **100** | **100** | **100** | **100** | **100** | **100** | **100** | **100** | **100** |
| | $N_B=6$ | **100** | **100** | **100** | **100** | **100** | **100** | **100** | **100** | **100** |
| $N_s=1000$ | $N_B=7$ | **100** | **100** | **100** | **100** | **100** | **100** | **100** | **100** | **100** |
| | $N_B=8$ | **100** | **100** | **100** | **100** | **100** | **100** | **100** | **100** | **100** |
| | $N_B=9$ | **100** | **100** | **100** | **100** | **100** | **100** | **100** | **100** | **100** |
| | $N_B=10$ | **100** | **100** | **100** | **100** | **100** | **100** | **100** | **100** | **100** |

Table 2 summarizes the success rate of ambiguity resolution as a function of the number of available BeiDou satellites, the number of particles, the number of integer candidates returned from the LAMBDA algorithm, and the quaternion bound of a prior attitude information. The number of satellites is artificially reduced in order to test the robustness of the proposed method against satellite-deprived environments. When the number of integer candidates is set to 15, the MCS approach achieves a 100% success rate no matter whether or not a priori attitude information is provided. However, when less candidates are used, the method only guarantees a 100% success rate with more than six or seven visible satellites. A decrease in the number of particles leads to decreased success rate, especially when less satellites are available and less candidates are used. A 100% success rate is always achieved at presence of a priori attitude information.

In addition to the MSC approach, the standard LAMBDA method and the constrained LAMBDA (C-LAMBDA) method [6,7,21] have also been implemented to provide comparison analysis. The LAMBDA method only achieves 100% success rate when more than eight satellites are available. The C-LAMBDA method incorporates the baseline length constraint and achieves 100% success rate when more than five satellites are available. In contrast, the MSC approach does not require code measurements and achieves 100% success rate even when only four satellites are available.

The computation time of the MSC method is summarized in Table 3. The algorithm has been implemented in the MATLAB software environment on a desktop computer. The computation time is mainly governed by the number of particles. Parallel computing could be employed to reduce the computation time for real-time applications.

**Table 3**. Average computation time of the Monte Carlo sampling approach for single-epoch processing.

| | $N_c = 5$ | $N_c = 10$ | $N_c = 15$ |
|---|---|---|---|
| $N_s = 100$ | 0.6 s | 0.7 s | 0.8 s |
| $N_s = 500$ | 2.1 s | 2.2 s | 2.4 s |
| $N_s = 1000$ | 4.1 s | 4.2 s | 4.3 s |

The attitude determination accuracy can be indicated from the standard deviations of attitude angles for static cases [22]. The three Euler angles, heading, elevation, and bank angles, are characterized by standard deviations of 0.15°, 0.42°, and 0.52°, respectively. The higher precision of the heading angle is due to the better observability of the GNSS system for the horizontal plane than for the vertical plane.

*4.2. Dynamic test*

A dynamic rotation experiment is designed to test the performance of the MSC approach for practical attitude determination. As seen in Fig. 3(a), three antennas were mounted on a horizontal rotating platform, which was surrounded by high buildings and trees. The antennas were connected to three low-cost K501 receivers, which were also manufactured by ComNav Technology of China. The nominal noise levels of code and phase measurements are 10 cm and 1 mm, respectively. The master antenna *M* and the first auxiliary antenna *A1* constitute the first baseline with a fixed length of 0.762 m. *M* and the second auxiliary antenna *A2* constitute the second baseline with a fixed length of 0.852 m. The local baseline coordinates are

$$\boldsymbol{F} = \begin{bmatrix} 0.762 & 0.380 \\ 0 & 0.763 \\ 0 & 0 \end{bmatrix} \tag{21}$$

A 1200 s observation data set were collected and analyzed offline. The time varying number of satellites tracked as well as the Position Dilution of Precision (PDOP) value are shown in Fig. 3(b).

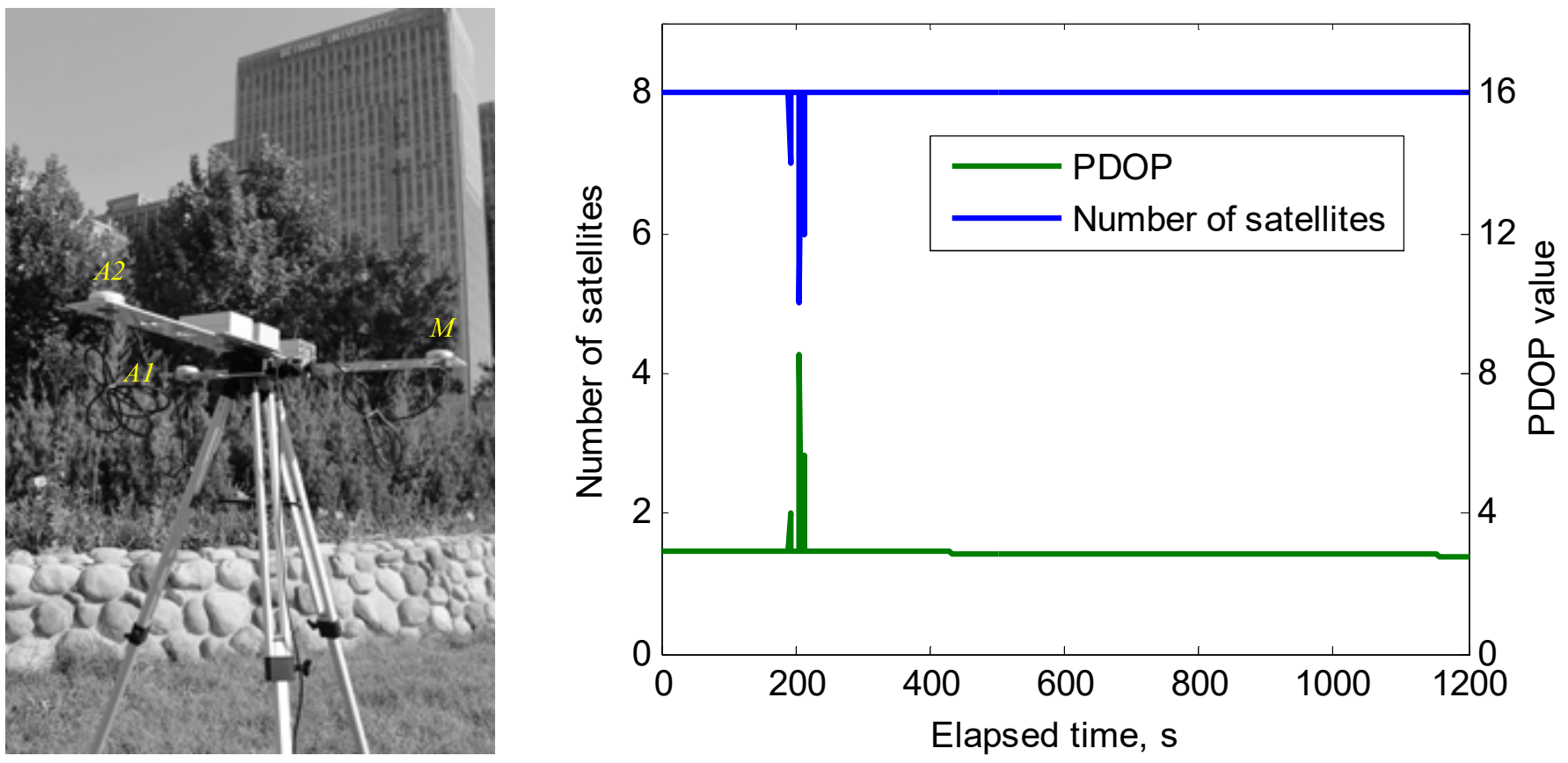

(a) Geometrical arrangement of the three antennas (b) Number of satellites tracked and PDOP values

Fig. 3. Dynamic BeiDou test: multipath urban environment (location: Beijing)

The success rate of ambiguity resolution for the dynamic case is presented in Table 4. The number of particles is set to 1000. The MCS approach achieves a 100% success rate for $N_c = 15$. However, the success rate decreases to 97.9 when the number of ambiguity candidates is set to 5 and no a priori attitude information is provided. The success rate of the dynamic case is slightly inferior to that of the static case. A major reason is that the baseline length is increased from 0.542 m to 0.852 m. The capability of using the baseline length to restrict the ambiguity probability space is weakened.

**Table 4**. Success rate (%) for Monte Carlo sampling approach in dynamic multipath environment. Rates of 100% are bolded.

| | $N_c = 5$ | | | $N_c = 10$ | | | $N_c = 15$ | | |
|---|---|---|---|---|---|---|---|---|---|
| Quaternion bound | ± 1 | ± 0.5 | ± 0.2 | ± 1 | ± 0.5 | ± 0.2 | ± 1 | ± 0.5 | ± 0.2 |
| Success rate | 97.9 | **100** | **100** | 99.6 | **100** | **100** | **100** | **100** | **100** |

## 5. Conclusion

The Monte Carlo sampling approach can achieve a 100% success rate of ambiguity resolution for instantaneous GNSS ultra-short baseline attitude determination. This new method is promising for attitude determination of size-limited vehicles under GNSS-challenging environments, e.g., with reduced number of available satellites, higher code noise or multipath.

## Acknowledgements

This research was supported by the National Natural Science Foundation of China through cooperative agreement No. 11002008 and has been funded in part by Ministry of Science and Technology of China through cooperative agreement No. 2014CB845303. The authors acknowledge the valuable remarks from the editor and anonymous reviewers.

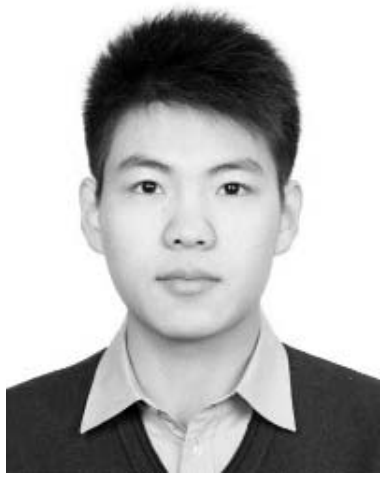

**Xiucong Sun** received his B.S. degree in aerospace engineering from Beihang University, Beijing, in 2010. He is a currently Ph.D. student at the school of astronautics, Beihang University. The focus of his research mainly lies in spacecraft orbit and attitude determination, GNSS navigation, gravity gradient based navigation.

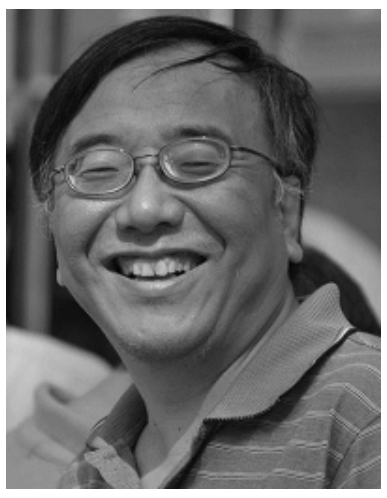

**Chao Han** received his M.S. and Ph.D. degrees in Applied Mechanics from Beihang University in 1985 and 1989, respectively. He is currently a professor at the school of astronautics, Beihang University. He is also an editorial board member of Chinese Journal of Aeronautics. The focus of his research activities lies in the area of spacecraft orbit and attitude dynamics, spacecraft guidance, navigation, and control.

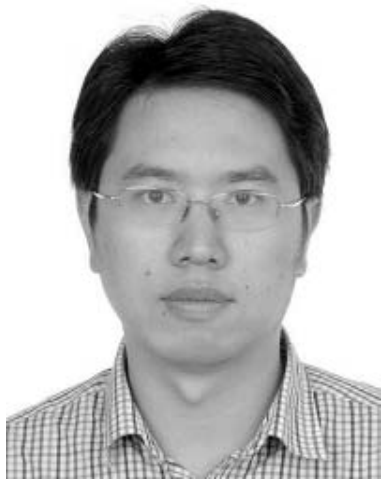

**Pei Chen** received his Ph.D. degree in aerospace engineering from Beihang University, Beijing, in 2008. He is currently an associate professor at the school of astronautics, Beihang University. His current research activities comprise spacecraft navigation, GNSS application, and astrodynamics and simulation.